# A Gravitational Shielding Based on *ZnS:Ag* Phosphor

Fran De Aquino

Maranhao State University,
Physics Department,
65058-970 S.Luis/MA, Brazil.
E-mail: deaquino@elo.com.br

## Abstract

It was shown that there is a practical possibility of gravity control on *electroluminescent* (EL) materials (physics/0109060). We present here a type Gravitational Shielding based on an EL *phosphor* namely zinc sulfide doped with silver (ZnS:Ag ) which can reduce the cost of the *Gravitational Motor* previously presented.

## Introduction

A recent experiment[1] using UV light on a *phosphorescent* material has detected a small reduction in the gravitational mass of the material. In a previous paper [2] we have explained the reported effect and presented a complete theory for the alterations of the gravitational field in *luminescent* (photo, electro, thermo and tribo) materials. We have obtained the following expressions of correlation between gravitational mass $m_g$ and inertial mass $m_i$ for particles under incident (or emitted) radiation (*fluorescent* or *phosphorescent* radiation):

$$m_g = m_i - 2\left\{\sqrt{1+\left\{\left(\frac{8\pi V}{c^3}\right)f^2 n_r^3\right\}^2} - 1\right\}m_i \qquad (1)$$

and,

$$m_g = m_i - 2\left\{\sqrt{1+\left\{\left(\frac{8\pi V}{c^3}\right)f^2 n_r^4\right\}^2} - 1\right\}m_i \qquad (2)$$

The Eq. (1) if the EL material is a *dielectric* (its electric conductivity $\sigma$ is such that $\sigma << \omega\varepsilon$, where $\omega = 2\pi f$; $f$ is the frequency of the light emitted from the EL material and $\varepsilon$ its electric permittivity). And Eq. (2) if the EL material is a *conductor* ($\sigma >> \omega\varepsilon$).

In Eqs. (1) and (2), $V$ is the *volume* of the particle( $m_i = \rho V$ where $\rho$ is matter density of the particle).

We see that the *index of refraction* $n_r$ of the EL material is a highly important factor in both equations, particularly in Eq. (2) where it appears with exponent 4.

In this paper we present a gravitational shielding based on an EL *phosphor* ($ZnS{:}Ag$) namely zinc sulfide doped with silver whose *index of refraction* is equal to 2.36.

The typical EL devices are consisted of light emitting ***phosphor*** sandwiched between two conductive electrodes. When an AC voltage is applied to the electrodes, the electric field causes the phosphor to rapidly charge and discharge, resulting in the emission of light during each cycle. The number of light pulses are determined by the magnitude of applied voltage, so, the brightness and color of the light emitted from the EL material will vary subject to the change of the operating voltage and frequency. For example, increasing the voltage increases brightness, whereas increasing the frequency of applied voltage will increase brightness and the *frequency* $f$ of the emitted light.

## 1. Gravitational Shielding

Let us consider a thin phosphor layer with chemical composition *ZnS:Ag*, index of refraction 2.36, and thickness $\xi$ on a metallic spherical shell with outer radius $R$.

When a specific alternating voltage $V_{AC}$ with frequency $f_{AC}$ is applied to the metallic spherical shell, the electric field causes the phosphor to rapidly charge and discharge, resulting in the emission of *blue* light ($f = 6.5\times10^{14}Hz$). Inside the phosphor this fluorescent radiation will fall upon the atoms. It can be shown that the probability of this radiation to reach the *nucleons* (protons and neutrons) of the atoms is practically null, thus we can assume that the radiation fall only upon the electrons, consequently changing, their gravitational masses. Under these circumstances the gravitational masses of the *electrons*, according to Eq. (2), will be given by:

$$m_{ge} = m_e - 2\left\{\sqrt{1+\left\{1.53\times10^8 R^2\xi\right\}^2} - 1\right\}m_e \quad (3)$$

and *the gravitational mass of the phosphor layer* will be

$$\begin{aligned} M_g &= n_1Z_1\left(m_{ge}+m_{gn}+m_{gp}\right)+ \\ &+n_2Z_2\left(m_{ge}+m_{gn}+m_{gp}\right)+ \\ &\ldots\ldots\ldots\ldots\ldots\ldots\ldots\ldots \\ &+n_nZ_n\left(m_{ge}+m_{gn}+m_{gp}\right)= \\ &=\left(n_1Z_1+n_2Z_2+\ldots+n_nZ_n\right)\left(m_{ge}+m_{gn}+m_{gp}\right)= \\ &=N\left(m_{ge}+m_{gn}+m_{gp}\right)= \\ &=N\left\{m_e-2\left\{\sqrt{1+\left\{1.53\times10^8R^2\xi\right\}^2}-1\right\}m_e+m_n+m_p\right\} \quad (4) \end{aligned}$$

$n_1$ is the number of atoms with atomic number $Z_1$ inside the phosphor layer, $n_2$ is the number of atoms with atomic number $Z_2$, etc.

When $M_g = 0$ the *phosphor coating* works as a *gravitational shielding*. This occurs if

$$m_e-2\left\{\sqrt{1+\left\{1.53\times10^8R^2\xi\right\}^2}-1\right\}m_e+m_n+m_p=0$$

i.e., if $R^2\xi = 1.19\times10^{-5}m^3$. For example if $\xi = 0.1mm$ then the external radius $R$ of the metallic spherical crust must be $R = 0.35m$.

It is important to note that we can vary the frequency $f$ of the emitted light to adjust the exact value of $M_g$. But, in the previous case where $R^2\xi = 1.19\times10^{-5}m^3$, if $f > 6.5\times10^{14}Hz$ the gravitational mass of the layer will be *negative*. This means that the weight of the phosphor layer will be *inverted*.

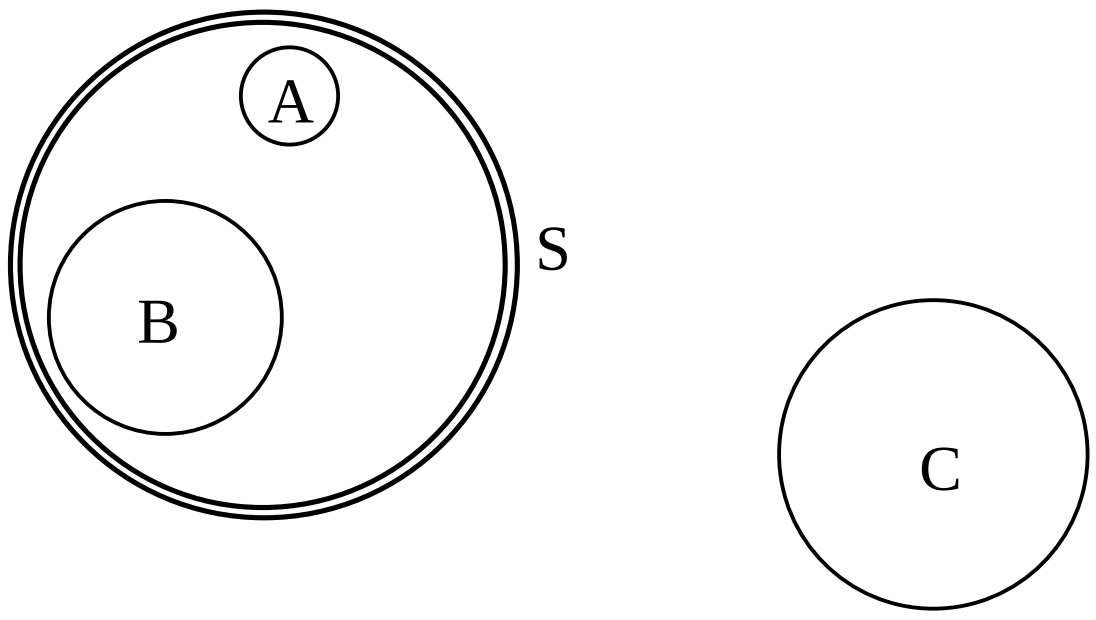

If the gravitational mass of the spherical shell S is equal to ZERO ($M_{gS} = 0$) then the gravitational forces between the bodies A,B,C and S are the following:

$$\vec{F}_{AB} = -\vec{F}_{BA} = -G\frac{m_A m_B}{r^2}\hat{\mu}$$

$\vec{F}_{AC} = -\vec{F}_{CA} = 0$
$\vec{F}_{BC} = -\vec{F}_{CB} = 0$ The interchange of gravitons between A and C, B and C is blockaded by S.

$\vec{F}_{BS} = -\vec{F}_{SB} = 0$
$\vec{F}_{CS} = -\vec{F}_{SC} = 0$ } $M_{gS}=0$

Fig.1-The Concept of Gravitational Shielding.

## 2. Gravitational Motor

From the technical point of view, the cost of the *Gravitational Motor* presented in recent papers[2,3] can be strongly reduced by means of the gravitational shielding above mentioned.

Let us consider the schematic diagram for the Gravitational Motor in Fig.2. Basically, it is similar to the previous design[2]. But the quantity of EL material in this model is very less than in the previous one. Consequently it will have a very smaller cost.

Now the idea isn't to invert the weight of the cylinders of EL material on the left side of the rotor but to *annul* the weight of *spheres of lead* by means of the gravitational shielding which coat them (see Fig.2).

In this design of the Gravitational Motor, the *average* mechanical *power* $P$ of the motor is given by:

$$P = T\omega = (2P_s\bar{r})\omega = (2M_{gs}g)\bar{r}\left(\frac{g}{\bar{r}}\right)^{\frac{1}{2}} =$$

$$= 2M_{is}\sqrt{g^3\bar{r}} \qquad (5)$$

where $\bar{r} = r - (R_0 + \Delta r)$ ; $R_0 = R + \xi$, (see Fig.2a) and $M_{is}$ is the *inertial mass* of *one* sphere.

Thus, if the radius of the rotor is $r = 0.35m$ *and* $R = 0.12m; \xi = 0.8mm;$ $\Delta r = 0.03m; M_{is} = 45kg$ we obtain

$$P \cong 1.2kw \cong 1.6HP$$

This is the power produced by *each group* of 6 spheres as shown in Fig.2a. Obviously additional similar groups can increase the total power.

An electric generator coupled at this motor can produce for one month an amount of electric energy $W$ given by

$$W = P.\Delta t = (1200w)(2.59\times10^6 s) =$$

$$= 3.1\times10^9 j \cong 861Kwh$$

We recall that the monthly residential consumption of electric energy of more than 99.7% of the Earth's residences is less than 800kwh.

## 3. Gravitational Spacecraft

In a previous paper[4] (Appendix B) we have presented a new concept of spacecraft and aerospace flight.

The new spacecraft, namely Gravitational Spacecraft, works by means of the gravity control.

The conception of the Gravitational Spacecraft is based on the fact that it is possible to make negative the gravitational mass of a body. Thus, we can imagine a spacecraft with *positive* $M_g^+$ and *negative* $M_g^-$ gravitational masses in such manner that its total gravitational mass $M_g$ is given by $M_g = M_g^+ + M_g^-$. The *negative* masse can be supplied by means of a system-G [4] ( Appendix A ) or by means of an EL material as the phosphor (*ZnS:Ag* ), in agreement with we have seen in the §1 of this paper.

In that case, to understand how it works, let us imagine an aircraft, which is totally coated with a layer of *ZnS:Ag*. When a appropriated voltage $V_{AC}$ with frequency $f_{AC}$ is applied on the metallic surface of the aircraft, the phosphor emits a fluorescent radiation (blue light ; $f = 6.5\times10^{14}Hz$ ) and consequently, as we have seen, inside the phosphor this fluorescent radiation will fall upon the electrons, consequently changing, their gravitational masses. Under these circumstances the gravitational masses of the *electrons*, according to Eq. (2), will be given by:

$$m_{ge} = m_e - 2\left\{\sqrt{1+\left\{\left(\frac{8\pi V}{c^3}\right)f^2 n_r^4\right\}^2} - 1\right\}m_e = \\ = m_e - Km_e \qquad (6)$$

where $V$ is the volume of the phosphor layer upon the aircraft.

If $K>>1$ the gravitational masses of the electrons will be given by

$$m_{ge} \cong -Km_e \qquad (7)$$

Consequently, from the Eq. (4) we can write the gravitational mass of the phosphor layer $M_{gs}$

$$M_{gs} = -NKm_e + N(m_p + m_n) \cong \\ \cong \frac{-N(Km_e) + N(m_p + m_n)}{m_{is}} m_{is} \cong \\ \cong \frac{-N(Km_e) + N(m_p + m_n)}{N(m_e + m_p + m_n)} m_{is} \qquad (8)$$

where $m_{is}$ is the *inertial* mass of the phosphor layer.

If $Km_e >> (m_p + m_n)$ the equation (8) reduces to

$$M_{gs} \cong \frac{-Km_e}{(m_e + m_p + m_n)} m_{is} \qquad (9)$$

Thus we can write the total gravitational mass $M_{ga}$ of the aircraft i.e.,

$$M_{ga} = M_{ga}^+ + M_{gs} \cong \\ \cong M_{ga}^+ - \frac{Km_e}{(m_e + m_p + m_n)} m_{is} \qquad (10)$$

$M_{ga}^+$ is the *positive* gravitational mass of the aircraft.

Consequently we can decrease the gravitational mass of the aircraft increasing the value of *K*. In practice will be possible to reduce the value of $M_{ga}$ up to some milligrams.

On the other hand, considering *the new expression for the inertial forces*, i.e., $\vec{F} = |M_g|\vec{a}$ deduced in a previous paper[4] , we can conclude that the aircraft can acquires an enormous acceleration $\vec{a} = \frac{\vec{F}}{|M_{ga}|}$ simply reducing its gravitational mass. In addition it is easy to see that the *inertial effects* on the aircraft will be strongly reduced due to the reduction of its gravitational mass. This is fundamentally a new concept of aerospace flight .

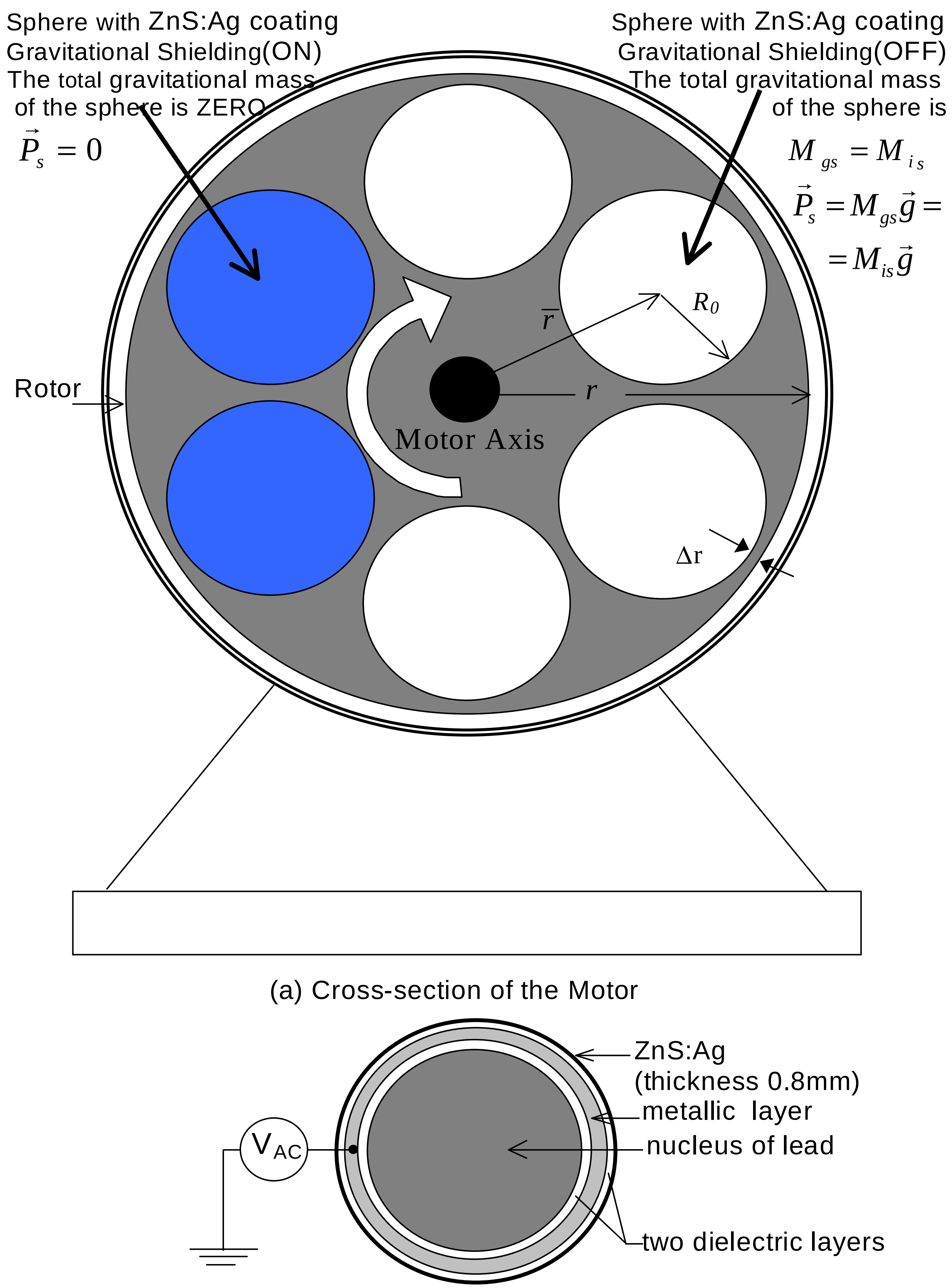

(a) Cross-section of the Motor

(b) Cross-section of the spheres

Fig.2 – The Gravitational Motor